\documentclass[
superscriptaddress,
twocolumn,
 amsmath,amssymb,
 aps,
prb,
a4paper,
dvipdfmx,
floatfix
]{revtex4-2}
\usepackage{graphicx,color}
\usepackage{dcolumn}
\usepackage{bm}
\usepackage{physics}
\usepackage[version=4]{mhchem}
\usepackage{gensymb}

\newcommand{\SLcat}{EtMe$_{3}$Sb }

\everymath{\displaystyle}
\usepackage{here}
\begin{document}
\title{Quantum Phase Transition of Organic Spin Liquid Tuned by Mixing Counterions}
\author{K. Ueda}
\affiliation{RIKEN, Condensed Molecular Materials Laboratory, Wako 351-0198, Japan}
\affiliation{National Institute of Technology, Anan College, Anan 774-0017, Japan}
\author{S. Fujiyama}
\email{fujiyama@riken.jp}
\affiliation{RIKEN, Condensed Molecular Materials Laboratory, Wako 351-0198, Japan}
\author{R. Kato}
\affiliation{RIKEN, Condensed Molecular Materials Laboratory, Wako 351-0198, Japan}
\date{\today}
\begin{abstract}
We found a plateau in the magnitude of the isolated magnetic moments as a function of the anisotropy of the transfer integral ($t'/t$) in the gapless quantum spin liquid (QSL) phase of an $S=1/2$ triangular lattice molecular solid \textit{X}[Pd(dmit)$_{2}$]$_{2}$, accomplished by a fine-tuning of $t'/t$ through the mixing of cations, \textit{X}. In contrast, the magnetic susceptibility at the lowest temperature in the QSL phase parametrized by $t'/t$ evinces an unconventional suppression approaching the quantum phase transition, implying significant critical fluctuations.
\end{abstract} 
\maketitle
\newpage


Antiferromagnetic ordered states can undergo quantum melting at $T=0$ K by changing physical parameters, resulting in a quantum phase transition (QPT) to a liquid-like state~\cite{Sachdev2011}. For example, in low-dimensional quantum spin systems or heavy fermion systems, the electronic coupling gives rise to a quantum-disordered state or a Fermi liquid~\cite{Chakravarty1989,Doniach1977}. 

Quantum spin liquid (QSL), driven by strong quantum fluctuation, is an example of the quantum disordered state. In two-dimensional (2D) systems, the geometrical frustration of antiferromagnetic correlation in triangular networks plays a vital role in destabilizing N\'{e}el antiferromagnetism and hosting the exotic quantum state~\cite{Anderson1973,Balents2010}. 

\begin{figure}[hbt]
  \includegraphics*[width=0.9\linewidth]{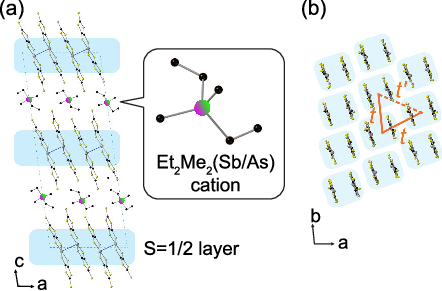}
  \caption{(a) Crystal structure of \textit{X}[Pd(dmit)$_{2}$]$_{2}$. The 2D [Pd(dmit)$_{2}$]$_{2}$ layer consists of an $S=1/2$ triangular lattice sandwiched between electronically closed shell cation layers. The bicolored atoms indicate As or Sb elements in the mixed cation crystals. (b) Top view of the $S=1/2$ magnetic layer. Secular transfer integrals $t$ and $t'$ are defined.}
  \label{fig:structure}
  \end{figure}

Molecular solids are known to realize QSL phenomena~\cite{Kanoda2011,Zhou2017}. Among them, EtMe$_{3}$Sb[Pd(dmit)$_{2}$]$_{2}$ (dmit = 1,3-dithiole-2-thione-4,5-dithiolate) consists of dimers formed by two Pd(dmit)$_{2}^{-1/2}$ molecules that host $S=1/2$ spins, forming a triangular network as shown in Fig.~\ref{fig:structure}. This 2D magnetic layer is sandwiched between electronically closed shell cation layers. The absence of a gap in the spin excitation spectrum has been widely observed in various experiments~\cite{Itou2008,Itou2010,Kato2014,Fujiyama2018,Pustogow2018,SYamashita2011,Watanabe2012}. However, there remains a discrepancy among experiments on whether a large Fermi surface of spinons is established~\cite{MYamashita2010,Bourgeois-Hope2019,Ni2019,Kato2022,Nomoto2022}. Recent \textit{ab initio} calculations suggest a Dirac-like linear dispersion~\cite{Ido2022}.

\begin{figure}[htb]
\centering
\includegraphics*[width=7.8cm]{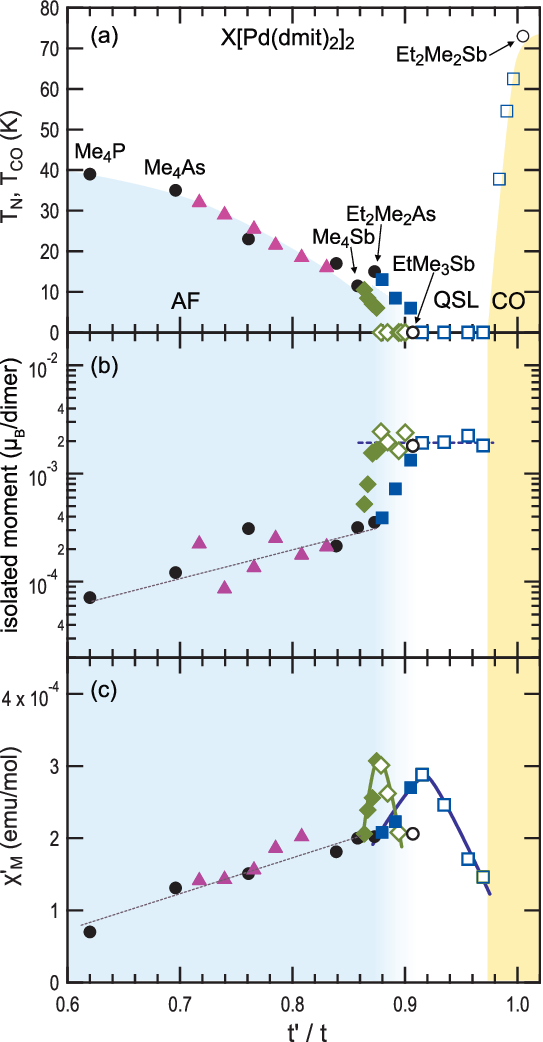}
\caption{(a) Magnetic ($T_N$) and charge-order ($T_\textrm{CO}$) temperatures of \textit{X}[Pd(dmit)$_{2}$]$_{2}$ as a function of $t'/t$. Solid circles show the $T_N$ for unmixed \textit{X}. $T_\textrm{CO}$ for \textit{X} = Et$_2$Me$_2$Sb, and the QSL of \textit{X} = EtMe$_3$Sb are shown as open circles.
The $T_N$ for \textit{X} = (Me$_{4}$As)$_{1-y}$(Me$_{4}$Sb)$_{y}$ are shown as violet triangles, while that for \textit{X} = (Me$_{4}$Sb)$_{1-y}$(EtMe$_3$Sb)$_{y}$ are depicted by solid green diamonds. $T_N$ for \textit{X} = (Et$_2$Me$_2$As)$_{1-y}$(Et$_2$Me$_2$Sb)$_{y}$ is shown as solid blue squares. Three blue open squares for $t'/t > 0.98$ show the $T_\textrm{CO}$. (b) Magnitudes of isolated magnetic moments, $M_\textrm{isolated}$. The $M_\textrm{isolated}$ demonstrate a constant value of $2 \times 10^{-3}\mu_{B}$ within the QSL phase. (c) $\chi'_{M}(=\chi_\textrm{max}-\chi(T=0))$ as a function of $t'/t$. $\chi'_{M}$ has a peak for each series of mixed cations.}
\label{fig:total}
\end{figure}

A notable feature of the \textit{X}[Pd(dmit)$_{2}$]$_{2}$ system is that we can manipulate the ground states to achieve antiferromagnetic ordered (AF), quantum spin liquid (QSL), and nonmagnetic charge-ordered (CO) states by cation substitution (\textit{X}$^{+}$)~\cite{Kato2014,Kanoda2011,Tsumuraya2013,Misawa2020}. So far, the AF or CO ordering temperatures, marked by circles in Fig.~\ref{fig:total}(a), have been parameterized by the anisotropy of the transfer integral, $t'/t$, which is determined by the lattice parameters for each \textit{X}. The decrease in $T_N$ with increasing $t'/t$, for $0.6 < t'/t <0.9$, suggests a specific value of $t'/t$ as a quantum critical point (QCP). The QSL state appears to be realized by strong quantum critical fluctuations near QCP since only one cation---\textit{X} = \SLcat---has been reported to exhibit QSL behavior~\cite{Tamura2009,Fujiyama2019,Li2019}, which supports the general phase diagram of QPT, claiming a unique gapless paramagnet at QCP~\cite{Sachdev2011,Chakravarty1989}. Although \textit{ab initio} calculations reproduce the decrease of $T_N$ with $t'/t$ ~\cite{Ido2022,Jacko2013,Kenny2021}, the phase transition from the antiferromagnetic phase to QSL is argued to originate from energy level crossing~\cite{Ido2022}.

There is no consensus on QSL stability. Interchain interactions easily destroy QSLs at low temperatures in one-dimensional spin chains, and small changes in control parameters can also lead to classical states in 2D triangular lattice network materials. Studies on \textit{X}[Pd(dmit)$_{2}$]$_{2}$, have found that only \textit{X} = \SLcat and its deuterated material show QSL behavior. 
As the deuteration can reduce $t'/t$ by an anticipated lattice contraction, references~\onlinecite{SYamashita2011} and~\onlinecite{Watanabe2012} argue that QSL exists not only at a QCP but also as a disordered phase. However, the controlled $t'/t$ width is minimal. Other effects, such as changes in lattice vibration due to deuteration, may mask it. To date, systematic material control across QPT keeping the Mott gap opened has not been achieved, and thus the quantum phenomena specific to QPT remain unexplored.

In this Letter, we demonstrate the magnetic susceptibility, $\chi$, of \textit{X}[Pd(dmit)$_{2}$]$_{2}$ with fine tuning of $t'/t$ realized by mixing the cations, \textit{X}. We obtained several QSL crystals with a finite range of $t'/t$ whose isolated magnetic moment remains constant, suggesting a QSL `phase' in the \textit{X}[Pd(dmit)$_{2}$]$_{2}$ system. In the vicinity of the QPT, we could not identify any sample undergoing an antiferromagnetic transition with $2 \le T_N < 6$ K due to the undetectable spin flop using polycrystalline samples, which prevented claiming a second-order transition. On the other hand, $\chi(T \rightarrow 0$ K) shows a clear minimum against $t'/t$, indicating unconventional critical fluctuations near the QPT.

\begin{figure}[htb]
\centering
\includegraphics*[width=0.9\linewidth]{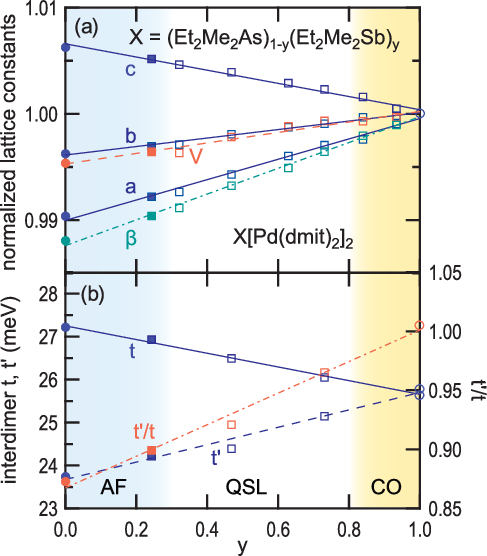}
\caption{Lattice parameters (a) and transfer integrals (b) of \textit{X} = (Et$_2$Me$_2$As)$_{1-y}$(Et$_2$Me$_2$Sb)$_{y}$. Closed (open) symbols show the parameters of the samples that undergo antiferromagnetic (nonmagnetic) states. Squares (circles) show parameters for mixed (unmixed) \textit{X}. (a) $a, b, c, \beta$ are the lattice parameters, and $V$ is the unit cell volume normalized by $y=1$. The space group is C2/c. (b) transfer integrals and the anisotropy evaluated by the extended H\"{u}ckel method. Parameters for \textit{X} = (Me$_4$Sb)$_{1-y}$(EtMe$_3$Sb)$_{y}$ and \ce{(Et2Me2As)_{1-y}(Et2Me2Sb)_y} are given in~\cite{Kato2012b}.}
\label{fig:lattice}
\end{figure}

We obtained \textit{X}[Pd(dmit)$_2$]$_2$ by slow air oxidation of \textit{X}$_2$[Pd(dmit)$_2$] in an acetone solution containing \textit{X}$^{+}$ and acetic acid at 5 \degree C. For example, in the case of \textit{X} = (Me$_4$Sb)$_{1-y}$(EtMe$_3$Sb)$_y$, the crystals were obtained by air oxidation of (Me$_4$Sb)$_2$[Pd(dmit)$_2$] and (EtMe$_3$Sb)$_2$[Pd(dmit)$_2$] in an acetone solution~\cite{Kato2012b}. The obtained crystals are mixed cations of two AFs, \textit{X} = Me$_4$As ($T_N = 36$ K) and \textit{X} = Me$_4$Sb ($T_N = 11.5$ K), \textit{X} = (Me$_{4}$As)$_{1-y}$(Me$_{4}$Sb)$_{y}$ ($0 \le y \le 1$), an AF, \textit{X} = Me$_4$Sb ($T_N = 11.5$ K) and the QSL, \textit{X} = EtMe$_3$Sb, \textit{X} = (Me$_4$Sb)$_{1-y}$(EtMe$_3$Sb)$_y$ ($0 \le y \le 1$), and an AF, \textit{X} = Et$_2$Me$_2$As ($T_N=15$ K) and a CO beyond the QSL state, \textit{X} = Et$_2$Me$_2$Sb, \textit{X} = (Et$_2$Me$_2$As)$_{1-y}$(Et$_2$Me$_2$Sb)$_{y}$ ($0 \le y \le 1$). The molar fraction of each cation in the mixed crystals was determined by electrospray ionization mass spectrometry (ESI-MS) with nitrobenzene solutions dissolving a single crystal. The molar fraction in the mixed crystal was finely controlled according to the initial molar fraction of each cation in the acetone solutions. As shown in Fig.~\ref{fig:lattice}, the lattice constants obtained by x ray diffraction (XRD) change continuously, from which we consider the two cations to form a homogeneous mixture in the crystal. The transfer integrals between the dimers were calculated by the extended H\"{u}ckel method with semi-empirical parameters based on the crystal structure obtained by single crystal XRD, from which we evaluated $t'/t$. We could tune the anisotropy of the transfer integrals, $t'/t$, with approximately the same steps, as shown in Fig.~\ref{fig:total}. It is to be noted that the variation of the intradimer transfer integrals, almost half of the onsite Coulomb repulsion, is limited within 2 \% to the substitions of cations, by which we consider the system maintains Mott insulating states~\cite{Kato2012,Tamura2004,Kanoda1997}. This is a notable synthetic character of \textit{X}\ce{[Pd(dmit)2]2}: chemical substitutions to $\kappa$-\ce{(BEDT-TTF)2Cu2(CN)3} QSL easily undergo metal-insulator transitions, and $t'/t$ no longer parameterizes the magnetic properties~\cite{Yoshida2019,Saito2021}. The magnetizations, $M$, of the mixed crystals were measured for $2 \le T \le 300$ K using 3--8~mg of samples using a superconducting quantum interference device (SQUID) under a magnetic field of 1~T. 

\begin{figure}[htb]
\centering
\includegraphics*[width=0.9\linewidth]{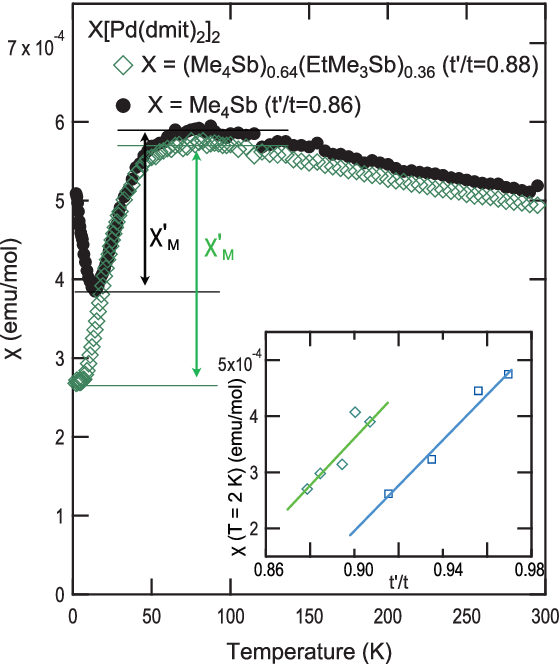}
\caption{Magnetic susceptibilities of \textit{X} = (Me$_4$Sb)$_{1-y}$(EtMe$_3$Sb)$_{y}$ $ (y=0.36, t'/t=0.88$, QSL) and \textit{X} = Me$_4$Sb $(t'/t=0.86, T_N=11.5$ K). We define $\chi'_M=\chi_\mathrm{max}-\chi_\mathrm{min}$, where $\chi_\mathrm{min}=\chi(T\rightarrow 0)$ K in the QSL phase. The inner core diamagnetism was evaluated by summing up the cation contribution value using Pascal's law and the previously reported value for \ce{[Pd(dmit)2]2}~\cite{Tamura2006}.
In the inset, we plot the susceptibilities at the lowest temperature ($T=2$ K) for various samples with different $t'/t$ in the \textit{X} = (Me$_4$Sb)$_{1-y}$(EtMe$_3$Sb)$_{y}$ (green diamonds) and (Et$_2$Me$_2$As)$_{1-y}$(Et$_2$Me$_2$Sb)$_{y}$ (blue squares) series. QPTs are located at $t'/t \approx 0.88$ for the former series and $t'/t\approx 0.92$ for the latter series.
}
\label{fig:definitionChiM}
\end{figure}

The temperature-dependent magnetic susceptibility, $\chi(T)$, of \textit{X}[Pd(dmit)$_{2}$]$_{2}$ shows a broad maximum ($\chi_\mathrm{max}$) at $T \approx 80$~K. It decreases at low temperatures, as shown in Fig.~\ref{fig:definitionChiM}. This trend is well reproduced by the high-temperature expansion (Pad\'{e} approximation) of the triangular lattice~\cite{Elstner1993,Tamura2002}. For AF materials, $\chi$ has a minimum at $T_N$. This temperature is consistent with the temperature at which the critical divergence of $1/T_1$ is observed by $^{13}$C NMR~\cite{Fujiyama2019}. 
$T_N$ decreases monotonically with $y$, as shown in Fig.~\ref{fig:total}~(a) for all three series of mixed cations, \textit{X} = (Me$_{4}$As)$_{1-y}$(Me$_{4}$Sb)$_{y}$ ($0\le y \le 1$), (Me$_{4}$Sb)$_{1-y}$(EtMe$_3$Sb)$_{y}$ ($0 \le y \le 0.21$), and (Et$_2$Me$_2$As)$_{1-y}$(Et$_2$Me$_2$Sb)$_{y}$ ($0 \le y \le 0.24$). The latter two series have different $T_N$ for the same $t'/t$ due to the error in estimating $t'/t$. Further substitutions suppress the AF states and induce QSLs; the crystals of $0.36 \le y \le 1$ in \textit{X} = (Me$_{4}$Sb)$_{1-y}$(EtMe$_3$Sb)$_{y}$ and $0.32 \le y $ in \textit{X} =  (Et$_2$Me$_2$As)$_{1-y}$(Et$_2$Me$_2$Sb)$_{y}$ show no evidence of magnetic ordering at $T \ge 2$~K. The systematic $t'/t$ dependence of $T_N$ shows that the precise tuning of $t'/t$ by synthesizing mixed cations is effective. Surprisingly, the observed QSL behavior for $0.32 \le y \le 0.73$ of (Et$_2$Me$_2$As)$_{1-y}$(Et$_2$Me$_2$Sb)$_{y}$ arises by mixing cations stabilizing the AF and CO states, manifesting the validity of $t'/t$ as a tuning parameter to describe the QPT.

This result gives a definite answer to the question of the stability of QSLs in \textit{X}[Pd(dmit)$_{2}$]$_{2}$. The QSL exists not only at a specific $t'/t$ critical point, but over a finite range of $t'/t$ values where QSL phases are stable. The QSL phase is consistent with the phase diagram of 2D quantum spin systems, where antiferromagnetic long-range ordered states reach quantum disordered states through QPTs by controlling physical parameters~\cite{Chakravarty1989}. On the other hand, we could not identify antiferromagnets with small $T_N$, below 6~K, in the obtained samples. Therefore, we cannot claim that the QPT from the AF to the QSL is a second-order transition.
\begin{figure}[htb]
\centering
\includegraphics*[width=0.85\linewidth]{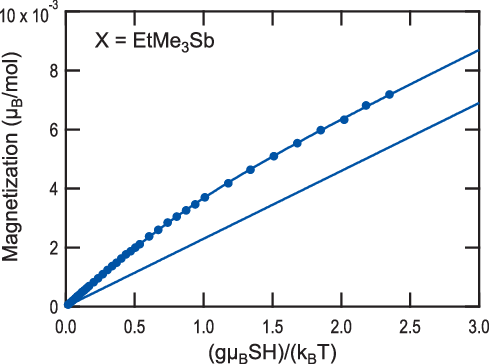}
\caption{Magnetization curve of \textit{X} = \ce{EtMe3Sb}. $\displaystyle M=\chi H + M_\mathrm{isolated}B_{J}(x)$ ($\displaystyle x=g \mu_B S H/k_{B} T$) fits the data. Linear plot denotes the paramagnetic component.}
\label{fig:magnetization}
\end{figure}

We obtained magnetization curves, $M(H)$, up to $H=7$~T at $T=2$~K. As shown in Fig.~\ref{fig:magnetization}, the magnetization is nonlinear in $H$, and the summation of a paramagnetic susceptibility and a Brillouin function, $\displaystyle M(H)=\chi H + M_\mathrm{isolated}B_{J}(x)$, fits well. Here, $M_\mathrm{isolated}$ is the magnitude of the isolated moment, and $B_{J}(x)$ is the Brillouin function parameterized by $\displaystyle x=g \mu_\text{B} S H/k_{B} T$, which saturates at $x \approx 1$. Here, $g, \mu_\text{B}$, and $S$ are the $g$ factor of the electron with $g = 2$, the Bohr magneton, and the spin quantum number $S = 1/2$, respectively. $M(H)$ indicates a small isolated moment $M_\mathrm{isolated} \approx 10^{-3 }\mu_\text{B}$ per molecular dimer. We observe the isolated moment in samples undergoing AF and the QSL. We plot $M_\mathrm{isolated}$ against $t'/t$ in Fig.~\ref{fig:total}(b). $M_\mathrm{isolated}$ increases by about a factor of five with the phase transition from the AF to the QSL phase and shows a plateau with $M_\mathrm{isolated} \approx 2 \times 10^{-3 }\mu_\text{B}$ in the QSL phase independent of $t'/t$.

$M_\mathrm{isolated}$ is tiny, which we may ignore as an extrinsic effect in conventional solid-state experiments. Nevertheless, since $M_\mathrm{isolated}$ of \textit{X} = EtMe$_3$Sb (a black circle in Fig.~\ref{fig:total}), and the two mixed cation series have comparable magnitudes, we can consider that $M_\mathrm{isolated}$ reflects an inherent characteristic of the QSLs. The fact that it is independent of $t'/t$ suggests a novel quantization mechanism that compensates for the isolated moment. Plateau magnetization can emerge under an external field in low-dimensional quantum spin systems. However, the magnitude of the magnetization plateau in $M(H)$ of \ce{NaYbO2}, a QSL candidate with triangular lattice, reaches e.g. 1/3 of the full polarization, which is considerably larger than the present case~\cite{Ranjith2019,Ding2019}. The cluster size of a recently investigated QSL candidate with charge modulation, 1\textit{T}-\ce{TaS2}, is $\approx10$, again incompatible with the small $ M_\mathrm{isolated} \approx 2 \times 10^{-3} \mu_\text{B}$~\cite{Ribak2017,Law2017}. Randomness in the antiferromagnetic correlation can destabilize the 120\degree N\'{e}el state and result in a QSL-like state~\cite{Fisher1994,Uematsu2021,Wu2019,Liu2018}. Here the magnetization is proportional to the number of orphan spins, which can be small. However, the reported calculations require extensive modulation in the antiferromagnetic correlation, which is implausible in our experiments because we did not perturb the $S=1/2$ 2D layer in the substitution of \textit{X}. The spin correlation must be long-range. In \textit{X}[Pd(dmit)$_{2}$]$_{2}$, the algebraic decay of the electronic correlation has been shown both experimentally and theoretically, and long-range correlations remain~\cite{Fujiyama2018,Ido2022,Kenny2021}, and this property can be closely related to the small value of $M_\mathrm{isolated}$ making the plateau.

Let us now discuss the reduction of $\chi(T)$ when $T<50$ K. We plot $\displaystyle \chi'_{M} \equiv \chi_\mathrm{max}-\chi_\mathrm{min}$ as a function of $t'/t$ in Fig.~\ref{fig:total}(c), where $\chi_\mathrm{min}$ is the minimum of $\chi(T)$ and $\chi_\mathrm{min} = \chi(T \rightarrow 0$ K) in the QSL phase. Unlike $M_\mathrm{isolated}$ in the QSL phase, which is insensitive to $t'/t$, $\chi'_M$ depends strongly on $t'/t$ and peaks at the boundary between the AF and the QSL phases. This peak in $\chi'_M$ indicates a critical point and we consider the peak of $\chi'_M$ as a phenomenon specific to the vicinity of the QPT. The $\chi'_M$ peak appears in a narrow range of $t'/t$, so it is unlikely that the antiferromagnetic coupling, $J \propto t^2$, changes significantly here. Therefore, $\chi_\mathrm{max}$ rarely depends on $t'/t$. The origin of the peak is that $\chi(T\rightarrow0$ K) depends strongly on $t'/t$ and achieves its minimum at the QPT. Indeed, $\chi(T=2)$ K in the QSL phase with different $t'/t$ in the \textit{X} = (Me$_4$Sb)$_{1-y}$(EtMe$_3$Sb)$_{y}$ and (Et$_2$Me$_2$As)$_{1-y}$(Et$_2$Me$_2$Sb)$_{y}$ series decrease approaching the QPT, as shown in the inset of Fig.~\ref{fig:definitionChiM}. This trend is consistent with other experiments. The transverse magnetic susceptibility, $\chi_\perp$, for \textit{X} = \SLcat and its deuterated material~\cite{Watanabe2012}, as well as the specific heat minus the $T^3$ term for \textit{X} = \SLcat and (Et$_2$Me$_2$As$_{0.25}$)(Et$_2$Me$_2$Sb$_{0.75}$)~\cite{Nomoto2022}, both exhibit reductions for reduced $t'/t$. When the electronic state is shifted away from the QCP, a smaller damping effect could cause steeper dispersions (a larger $\omega/k$ relation) for the magnetic excitation, resulting in a smaller density of states and $\chi$. The observed suppression of $\chi$ near the QPT contradicts this naive expectation.

Finally, we mention the relationship between the QPT and superconductivity (SC)---a macroscopic quantum phenomenon. So far, only \textit{X}[Pd(dmit)$_{2}$]$_{2}$ with AF ground states at ambient pressure exhibits superconductivity under pressure, but \textit{X} = EtMe$_{3}$Sb does not undergo the SC transition~\cite{Kato2014}. Applying physical pressure enhances the kinetic energy of the electrons and also causes a significant change in $t'/t$. For example, in \textit{X} = Et$_2$Me$_2$P, which exhibits SC at 6.9~kbar at $T_c=4$~K, the increased pressure reduces $t'/t$ as shown in Table~\ref{tab:aniso}~\cite{Yamaura2004}. The physical pressure can relocate $t'/t$ away from the QPT, which implies a limitation of a simplified scenario for quantum fluctuations near the QCP to drive SC. Despite both the material design to tune $t'/t$ near the QPT and the physical pressure causing a damping of the spin correlation, there are still differences between the two effects. For example, the former is unrelated to the Mott transition. There persists the non-trivial issue of establishing a route from QSL to SC that relies on a category of the QSL state.
\begin{table}[H] 
 \caption{Anisotropy of the transfer integrals, $t'/t$, of \textit{X} = Et$_2$Me$_2$P under pressure~\cite{Yamaura2004}.\label{tab:aniso}}
 \begin{ruledtabular}
 \begin{tabular}{c c  c  c}
 Pressure (kbar) & $1\times 10^{-3}$ & $7.2$  & $16.8$ \\
$t'/t$ & 0.841 & 0.705 & 0.630 \\
 \end{tabular}
 \end{ruledtabular}
 \end{table}
 
In summary, we have exhibited the magnetic susceptibility of \textit{X}[Pd(dmit)$_{2}$]$_{2}$ with the precise tuning of $t'/t$ achieved by mixing cations, \textit{X}. We obtained several QSL crystals, even mixing cations stabilizing the antiferromagnetic and charge-ordered states, with a finite range of $t'/t$ indicating a QSL `phase' in the \textit{X}[Pd(dmit)$_{2}$]$_{2}$ system. We found a plateau phenomenon in the isolated magnetic moments with $M_\mathrm{isolated} \approx 2 \times 10^{-3}\mu_\text{B}$ in the QSL phase, the origin of which remains unresolved. We could not observe a second-order AF-QSL phase transition due to the undetectable magnetic anisotropy, which prevented us from obtaining AF samples with $2 \le T_N < 6$ K. On the other hand, $\chi'_M$ shows a distinct peak at the AF-QSL phase boundary due to the reduction of $\chi(T \rightarrow 0$) K near the QPT, which we consider a novel manifestation of the quantum critical fluctuation.

\begin{acknowledgments}
We are grateful to H.~Seo, H.~Maebashi, H.~Matsuura, Y~Otsuka, and T.~Momoi for fruitful discussions. This work was supported by Grants-in-Aid for Scientific Research (20K03870, 21K03426, 21K05001) from JSPS.
\end{acknowledgments}


%
  
\end{document}